# 2D THz spectroscopic investigation of ballistic conduction-band electron dynamics in InSb

S. HOUVER[*] AND L. HUBER, M. SAVOINI, E. ABREU, S. L. JOHNSON

*Institute for Quantum Electronics, ETH Zürich, 8093 Zurich, Switzerland*
*\*shouver@phys.ethz.ch*

**Abstract:** Using reflective cross-polarized 2D THz time-domain spectroscopy in the range of 1-12 THz, we follow the trajectory of the out-of-equilibrium electron population in the low-bandgap semiconductor InSb. The 2D THz spectra show a set of distinct features at combinations of the plasma-edge and vibration frequencies. Using finite difference time domain simulations combined with a tight binding model of the band structure, we assign these features to electronic nonlinearities and show that the nonlinear response in the first picoseconds is dominated by coherent ballistic motion of the electrons. We demonstrate that this technique can be used to investigate the landscape of the band curvature near the Γ-point as illustrated by the observation of anisotropy in the (100)-plane.

## 1. Introduction

The development of ultrafast laser systems and light conversion techniques has enabled the generation of few-cycle terahertz (THz) pulses with high electric fields and has enabled a new range of nonlinear optical studies. For carrier dynamics in semiconductors in particular, a variety of high-field effects have been demonstrated using ultrafast THz spectroscopy. These include interband tunneling in GaAs [1], impact ionization and intervalley scattering [2-5], Bloch oscillations in GaSe [6] and ballistic transport of electrons in GaAs and InGaAs [7, 8]. For the latter, carriers are freely accelerated by electric fields along the conduction band, leading to a coherent optical response which contains information about the band structure of the material [7, 8].

THz multidimensional spectroscopy provides additional spectral information compared to conventional ultrafast THz spectroscopy techniques. By probing the nonlinear response of the system along two independent frequency components, it enables studying the coupling between different degrees of freedom (electrons, lattice, etc.) and also purely electronic phenomena such as the ballistic transport of electrons. In the study of condensed matter systems, 2D THz spectroscopy has so far been used in the time domain to investigate different types of excitations and their couplings, such as the correlations of electronic and lattice excitations in GaAs/AlGaAs quantum wells [9], two-phonon coherences in bulk InSb [10] and magnon coherence and correlations in YFeO$_3$ [11]. For most of these studies relatively narrowband THz radiation is tuned to be close to resonance with intrinsic material frequencies (intersubband resonances or phonon/magnon resonances for example). 2D THz spectroscopy has also been implemented to investigate nonlinear dynamics of electrons in various low-bandgap systems [12-15]. In many materials there are a wide range of frequencies in the THz range that are relevant for studies of electron dynamics and coupling, making broadband THz pulses combined with 2D THz spectroscopy a highly attractive technical goal.

In this letter, we present cross-polarized 2D THz spectroscopy measurements over a broad 1-12 THz range to investigate the time-resolved electronic band nonlinearities of the low-bandgap semiconductor InSb during the first few picoseconds after excitation by a strong, nearly single-cycle THz field. With this method and precise controls over the field polarization, we can perform a complete parity analysis and thus are able to distinguish nonlinear contributions from physically distinct mechanisms. Although electron scattering effects are known to be sub-100 fs, we show that the coherent ballistic motion of field-driven electrons dominates the nonlinear response in the first few picoseconds, as previously demonstrated in GaAs films [8]. Our experimental observations are supported by finite-difference time-domain (FDTD) simulations [16] of the ballistic response of InSb to the THz electric fields. These enable us to identify the conduction band curvature characteristics that give rise to the observed nonlinearities.

## 2. Experimental methods

In order to create the THz pulses used in our measurements, we start with an amplified femtosecond laser system. The output of a Ti:Sapphire amplifier ($\lambda$= 800 nm, 100-fs pulse, repetition rate: 1 kHz) seeds two three-stage optical-parametric-amplifiers (OPAs) resulting in two output beams tuned at 1.3 μm and 1.5 μm, with 1 mJ/pulse and 1.5 mJ/pulse respectively. These are used to generate broadband THz pulses from two separate sources: a two-color plasma source [17] providing up to 100 kV/cm peak field (amplitude) at the sample position with a 1-12 THz bandwidth and a source based on optical rectification (OR) in an organic crystal (DSTMS) [18] providing up to 250 kV/cm peak field with 1.5-4.5 THz spectral content. Two independent choppers on the plasma source and OR source paths are set to allow transmission of the light pulses at a quarter and half of the repetition rate of the laser, respectively, in order to excite the sample with either the "probe" field $E_{plasma}$, the "pump" $E_{OR}$ or the combined fields $E_{OR+p}$. The inset of figure 1a shows a schematic of the temporal pulse sequence.

A key feature of this experiment is the careful control of the two THz beam polarizations to obtain cross-polarized electric fields to excite the sample. The THz beam from plasma generation is horizontally polarized, which is maintained until it interacts with the sample. The THz beam generated from OR is also initially horizontally polarized, but is rotated to vertical polarization using a sequence of 2 wire-grid polarizers (WGs). In this scheme the first polarizer is oriented in order to pass electric fields polarized at an angle $\theta$ with respect to the horizontal, and the second is oriented to allow only vertically polarized electric fields to pass. By rotating $\theta$ between values of -45° and +45°, we can precisely control both the amplitude and sign of the electric field pulses transmitted by the second WG. The two pulses are combined on the sample at normal incidence using another WG polarizer that transmits the horizontally-polarized plasma-generated pulse and reflects the vertically-polarized OR-generated pulse, as shown in Fig. 1a. The horizontally-polarized plasma-generated pulse reflected by the sample gets transmitted through the combining WG and then is partly reflected on a Si beamsplitter towards the detection cell. There, the THz field is detected using broadband Air-Biased Coherent Detection (ABCD) [19] with a split-off sampling beam at 1.3 μm, mostly sensitive to horizontally polarized beams. As the cross-polarizations are well-defined, the part of the OR-generated pulse which is reflected by the sample is mostly reflected back to the generation crystal at the combining WG so that despite the collinear incidence geometry, the detection is nearly insensitive to the OR-generated field.

Examples of the time- and spectral-dependence of the field $E_{plasma}$ generated by the plasma source (red) and that of the field $E_{OR}$ generated by the OR source (green) are shown in figures 1b and 1c. These traces were measured using the broadband ABCD method at the detection position. To obtain accurate electric field measurements at the sample position, we temporarily replaced the sample by electro-optic sampling detection, using a 100 μm-thick GaP crystal. Due to its limited spectral sensitivity towards higher frequencies, this measure gives a lower estimate of the actual field strength for $E_{plasma}$. In the following, the indicated peak field amplitudes refer to these EO-sampling measurements at the sample position. The delays of the sampling infrared beam for ABCD and the pumping infrared beam for the OR source are independently controlled, and in the following will be referred to as "detection delay t" and "excitation delay τ", respectively. For the 2D measurements, both delays are scanned and the nonlinear signal $E_{NL}$, defined as $E_{NL} = E_{OR+p} - E_{plasma} - E_{OR}$, can thus be extracted, for every delay t and τ. The detection specificity to horizontally-polarized fields is further enhanced by placing an additional WG polarizer before the detection cell.

For 1D reflectivity measurements, the setup is used as a broadband time-domain spectrometer (TDS) with the plasma field only. In order to get a reference signal at the exact position of the sample, an intense 1500-nm-laser pulse illuminates the semiconducting sample, inducing a transient state with a plasma frequency far beyond the observed frequency range. Using this *photo-screening* method, we briefly create a mirror-like reference at the exact position of the sample. The sample equilibrium reflectivity can be determined by comparing the reference with the semiconducting state response measured with the following laser pulse 1 ms later [20].

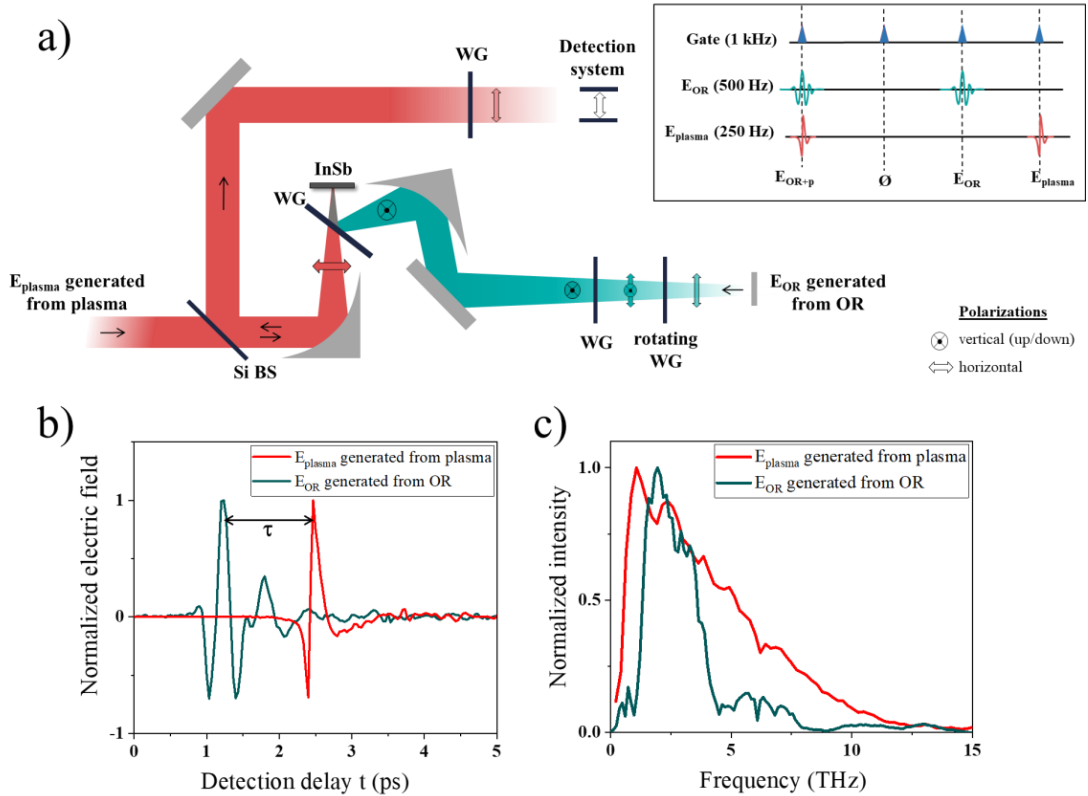

Fig. 1. a) Measurement geometry of reflective 2D THz spectroscopy on InSb at normal incidence. The horizontally-polarized plasma field, $E_{plasma}$, is transmitted back and forth through the wire-grid polarizer (WG), while the vertically-polarized field from OR in the organic crystal, $E_{OR}$, is mostly reflected. The nonlinear response is retrieved using the chopping scheme depicted in the inset. b) $E_{plasma}$ (red) and $E_{OR}$ (green) normalized electric fields, measured with broadband ABCD method. c) Corresponding $E_{plasma}$ (red) and $E_{OR}$ (green) normalized spectra.

3. Results

Figure 2a shows the reflectivity amplitude (black circles) and phase (green circles) measured on the (100)-cut bulk nominally-undoped InSb sample by TDS using a strongly attenuated broadband plasma source and the photo-screening method to reference the data. The reflectivity data can be fitted using the combined Drude-Lorentz model for the dielectric function, given as:

$$\varepsilon = \varepsilon_\infty \left( 1 + \frac{\omega_{LO}^2 - \omega_{TO}^2}{\omega_{LO}^2 - \omega^2 - i\gamma_{ph}\omega} - \frac{\Omega_p^2}{\omega^2 + i\gamma_p\omega} \right) \qquad (1)$$

where $\omega_{TO}$ and $\omega_{LO}$ are transverse (TO) and longitudinal optical (LO) phonon frequencies respectively, $\gamma_{ph}$ is the corresponding phonon damping, $\Omega_p$ the plasma frequency of Drude carriers with damping $\gamma_p$, and $\varepsilon_\infty$ represents the high frequency dielectric function. Fitting the complex reflectivity, we extract all parameters: the plasma-edge frequency $\nu_p = \Omega_p/(2\pi\sqrt{\varepsilon_\infty}) = 2.2\,\text{THz}$, TO and LO-phonon frequencies $\nu_{TO} = 5.3$ THz and $\nu_{LO} = 5.7$ THz, $\varepsilon_\infty = 15.8$ and the damping $\gamma_p = 0.3$ THz and $\gamma_{ph} = 0.6$ THz. All are in good agreement with previously reported data [21], despite common variations in growth-process that affects doping and thus the Drude coefficients.

Figure 2b shows the 2D temporal trace of the nonlinear signal acquired on InSb varying detection delay $t$ and excitation delay $\tau$ with peak electric fields of $E_{plasma} \sim 25$ kV/cm and $E_{OR} \sim 65$ kV/cm. The field $E_{plasma}$ has a constant phase at a given detection delay $t$, while $E_{OR}$ evolves following the diagonal with a delay $t + \tau = 0$ (dashed line). For delays $t + \tau > 1.5\,\text{ps}$, the nonlinear signal shows a vertical structure almost completely independent of the excitation delay, which can be understood as a persistent change of the reflectivity of $E_{plasma}$ induced by $E_{OR}$. For InSb, these changes can be attributed to carrier multiplication induced by impact ionization [3]. In the first picosecond, where both THz pulses temporally overlap, the nonlinear signal shows a clear dependence on both detection delay $t$ and excitation delay $\tau$. This nonlinear contribution amounts to about 15% of the incoming peak field $E_{plasma}$.

The different signal modulations observed in the time domain transform into distinct peaks in frequency space, as shown by the 2D spectrum presented in figure 2c. We can assign these features to combinations of plasma-edge frequency $\nu_p$ and LO phonon frequency $\nu_{LO}$ that were measured with TDS (fig. 2a). We note here that the peaks do not appear exactly at the plasma-edge frequency previously measured (2.2 THz), but rather at the maximum of the phase shift of the reflectivity (1.9 THz). Expressed as ($\nu_t$, $\nu_\tau$), the main observed peaks here are ($\nu_p$, 0), ($\nu_p$, $2\nu_p$), ($\nu_{LO}$, 0) and ($\nu_{LO}$, $2\nu_p$).

In order to distinguish the various contributions to the nonlinear response in the 2D frequency domain and to separate coherent and incoherent contributions, we can invert the polarity of $E_{OR}$ (phase shift of 180°) and then extract the *odd* ($E^-$) and the *even* ($E^+$) components of the nonlinear signal. A similar approach was followed for 1D measurements by Ho et al. [22]. For 2D experiments, odd and even components are a combination of 5 different fields, defined as: $E^\pm = \left(E_{OR+p} - E_{plasma} - E_{OR}\right) \pm \left(E_{\overline{OR}+p} - E_{plasma} - E_{\overline{OR}}\right)$ where $\overline{OR}$ indicates the inverted polarity of the OR field.

The even component $E^+$ contains nonlinear terms with even orders of the OR field, including incoherent contributions from carrier multiplication proportional to the field intensity. The odd signal, conversely, contains nonlinear terms with odd orders of the OR field and is then expected to show only coherent responses that have a fixed phase relation to $E_{OR}$. Because our detection scheme does not completely eliminate contributions from $E_{OR}$, we might expect some contributions in the "odd" signal arising from an influence of $E_{plasma}$ on $E_{OR}$. These contributions are, however, very close to the noise floor of our measurement.

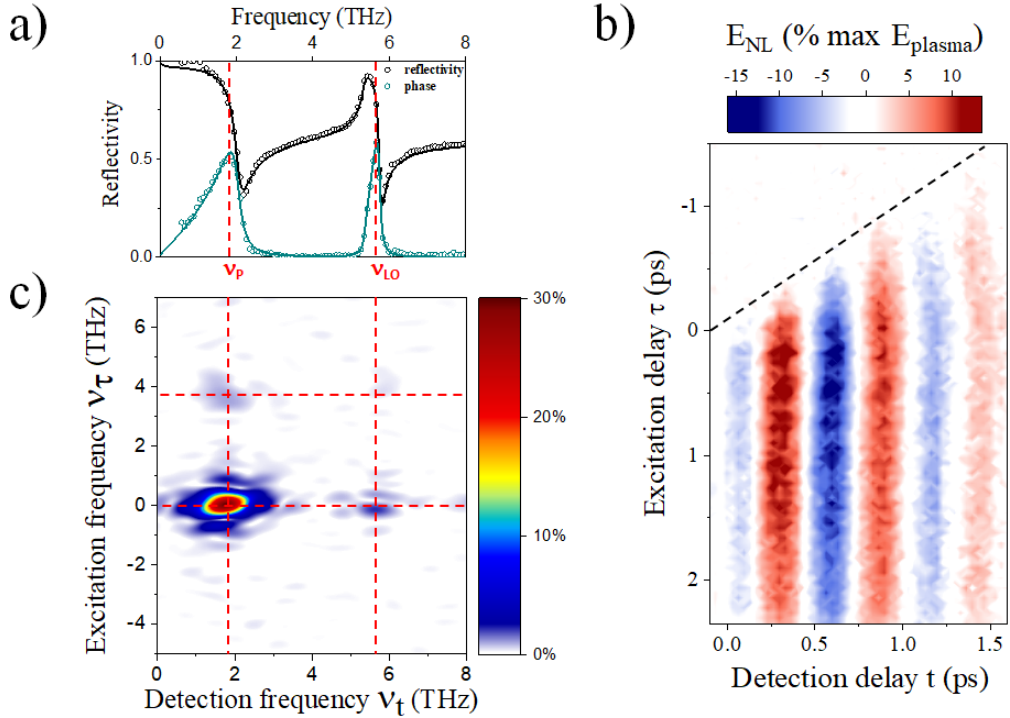

Fig. 2. a) Reflectivity amplitude (black circles) and phase (green circles) of InSb measured by TDS. Solid lines show fits to the complex reflectivity data, based on equation (1). Dashed red lines indicate the maxima in the phase-shift of plasma-edge and LO phonon frequencies at 1.9 THz and 5.7 THz, respectively. b) Nonlinear response $E_{NL}(t,\tau)$ of InSb measured with polarizations $E_{plasma}$ and $E_{OR}$ along (010) and (001), respectively, as a function of detection delay t and excitation delay τ, for $E_{plasma}$ ~ 25 kV/cm and $E_{OR}$ ~ 65 kV/cm. The color scale

is normalized to the maximum of $E_{plasma}$ 2D measurement. c) 2D spectrum obtained from the Fourier transform of $E_{NL}(t,\tau)$, plotted as a function of detection frequency $\nu_t$ and excitation frequency $\nu_\tau$. Values are given with respect to maximum of $E_{plasma}$ 2D spectrum.

In order to investigate any possible anisotropy in the nonlinear response of the sample, we studied the dependence of the 2D spectrum on the angle between the THz electric field and a reference axis of the InSb sample. The polarizations of $E_{plasma}$ and $E_{OR}$ are kept orthogonal, as described in figure 1a, but the sample is rotated so that the THz electric fields excite different directions. We define the angle $\varphi$ as the angle between the $E_{plasma}$ polarization and the in-plane InSb axis (010) which in reciprocal space corresponds to the angle between $E_{plasma}$ and the $\Gamma \rightarrow X$ direction in the Brillouin zone.

Figures 3a and 3b show odd and even 2D spectra for two sets of measurements. Two different $\varphi$ angles were acquired per set, $\varphi = 45°$ and $\varphi = 22.5°$ for fig. 3a and $\varphi = 45°$ and $\varphi = 67.5°$ for fig. 3b. In both cases, we see that odd and even spectra present different features. The even signal $E^+$ is similar for all angles and presents the main characteristics previously pointed out in fig. 2c, with the strongest peak at $(\nu_p,0)$ (relative intensity of 30%). The odd signal $E^-$ shows a weak but well-defined feature on the diagonal at $(\nu_p, \nu_p)$ (relative intensity near 2%). This feature appears for $\varphi$ angles of 22.5° and 67.5° (on figures 3a and 3b respectively) but is at the noise level for $\varphi = 45°$ (in both figures) and thus indicates an anisotropic response.

## 4. Simulations

To better understand the features of the 2D spectra observed experimentally, we model the response of the InSb bulk sample in the temporal overlap region beyond the perturbation limit. Here we model the electronic motion in the first few picoseconds as a coherent wave packet motion of conduction band electrons [8, 23]. Indeed, these systems have a long electronic coherence time [8] and the probability of tunneling from the valence to the conduction band is reduced by the low joint density of states at the $\Gamma$-point. In the following, we will hence neglect electronic decoherence and approximate the dynamics as those of carriers moving in a two-dimensional conduction band. Our model therefore describes the ballistic motion of electrons, driven by THz electric fields in the conduction band but does not include interband tunneling or impact ionization or any scattering channels apart from those in the Drude-Lorentz model.

To implement the physics of electronic motion and the interaction of polar phonons with few-cycle THz pulses, we solve the time dependent Maxwell's equations in the presence of an interacting medium, using the FDTD method [16, 23]. This algorithm, together with a description of ballistic motion of conduction band carriers for a polar semiconductor in response to electric fields, enables modeling of the response in the temporal overlap region beyond the perturbation limit. To allow for arbitrary crystal orientations and two dimensional carrier motion, we include the full band structure, obtained from tight binding calculations [24], in our FDTD simulations. In the absence of a magnetic field the spin-split conduction bands are equally populated and we therefore approximate the conduction band as an average of the spin-split bands. Following the work by Yu et al. [23], the local electric displacement in the semiconductor is given by:

$$D(z) = \varepsilon_0 \varepsilon_\infty E_{THz}(z) + P_{ph}(z) + P_e(z)$$

where $E_{THz}$ is the driving THz electric field, $P_{ph}$ and $P_e$ correspond to the phonon and electron polarization densities respectively and the dielectric background is given by the scalar $\varepsilon_\infty$. The phonon polarization density $P_{ph}$ is implemented as a Lorentz oscillator with given strength, damping and frequencies, previously introduced in equation (1). The temporal evolution of the electronic polarization density $P_e$ is then given by:

$$\frac{dP_e}{dt} = -e n_e v_g(k(t))$$

with $n_e$ being the carrier concentration and $v_g(k)$ the group velocity of an electron wave packet of wave vector $k(t)$. The temporal evolution of wave vector $k$ is computed from the electron response to the THz electric field, governed by the equation of motion:

$$\hbar \frac{dk}{dt} + \gamma_p \hbar k = -e E_{THz}(z,t)$$

where $\gamma_p$ is the electron scattering rate. For each $k(t)$ position and using a semi-classical description, the group velocity $v_g(k)$ is given by:

$$v_g(k) = \frac{1}{\hbar} \frac{\partial E_{TB}(k)}{\partial k}$$

where $E_{TB}(k)$ is the conduction band dispersion relation obtained from tight biding calculations.

Although a change in scattering rate is known to occur for higher energy electrons, we keep the damping $\gamma_p$ at its equilibrium value and do the same for the carrier density $n_e$. The equilibrium values, listed above, were found by fitting the measured linear reflectivity shown in figure 2a, so that the simulation has no free parameters.

Using the FDTD method with the previously described assumptions, we model the response of bulk (100)-InSb excited by two THz electric fields. The THz electric field waveforms used in the simulations come from a fit to the experimental plasma and OR temporal traces and we set $E_{plasma} = 35$ kV/cm and $E_{OR} = 65$ kV/cm. The simulated linear response to the plasma field excitation alone (not shown) is in very good agreement with the measured reflectivity spectrum presented in fig. 2a. Under the excitation of both $E_{OR}$ and $E_{plasma}$, the nonlinear signal obtained in the temporal overlap is about 15% of the incoming $E_{plasma}$ field, comparable to the experimental nonlinear signal presented in figure 2b.

By Fourier transforming the temporal responses, we obtain simulated odd and even spectra for different configurations. The simulated spectra for $\varphi = 22.5°$ and for $\varphi = 67.5°$ are included in figure 3 for comparison with the experimental data. As expected from the fourfold symmetry of the potential in this plane, both angles show very similar results in the simulations. The even spectrum shows the features already pointed out in fig 2c, at $(\nu_p, 0)$, $(\nu_p, 2\nu_p)$, $(\nu_{LO}, 0)$ and $(\nu_{LO}, 2\nu_p)$. The main peak $(\nu_p, 0)$ intensity is close to 50% of the $E_{plasma}$ spectrum maximum. The odd spectrum shows mainly the peak $(\nu_p, \nu_p)$, with a relative intensity near 2%. For $\varphi = 45°$ (not shown), the simulated spectra lead to a similar even signal and to a vanishing odd signal.

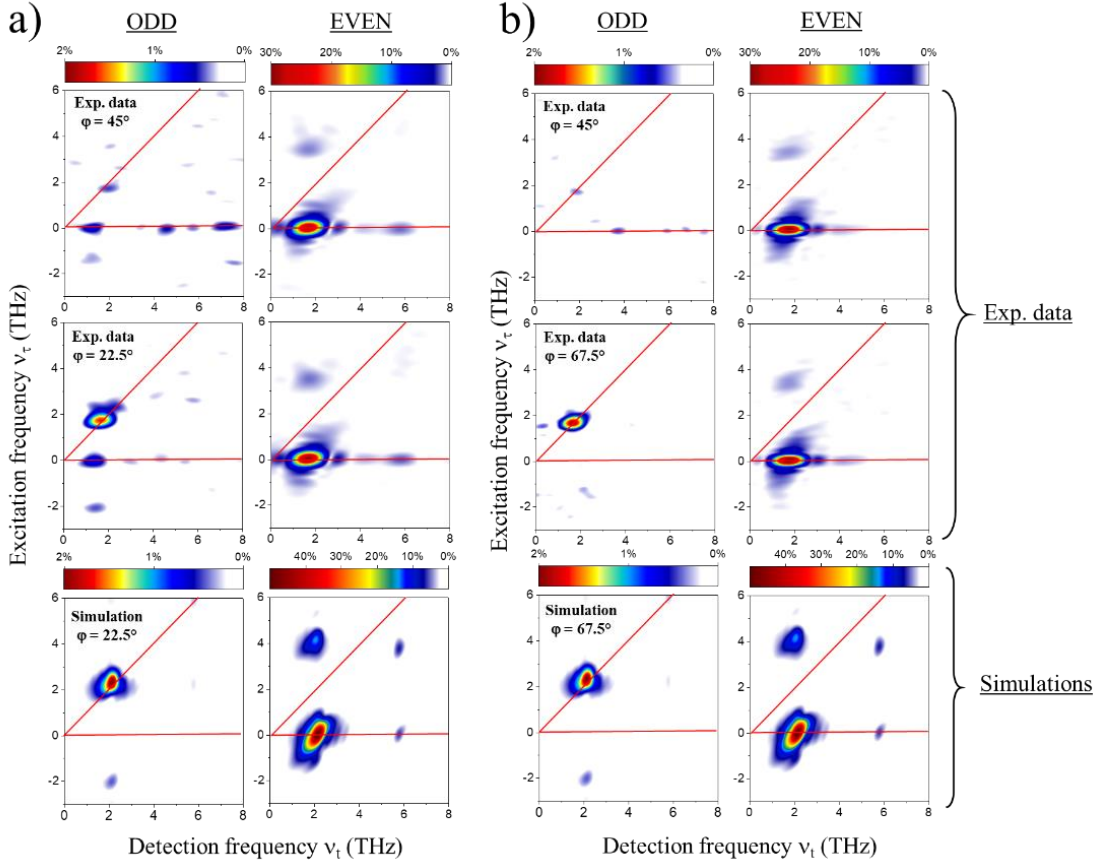

Fig. 3. a) Odd and even parity spectra of InSb for $E_{plasma} \sim 35$ kV/cm and $E_{OR} \sim 65$ kV/cm. The first two sets show the experimental results from the same measurement at two different angles, $\varphi = 45°$ and $\varphi = 22.5°$. The last set shows simulated spectra (FDTD) for an angle $\varphi = 22.5°$. b) Odd and even parity spectra for similar electric fields. First two sets: experimental results for the angles $\varphi = 45°$ and $\varphi = 67.5°$; last set: simulated spectra at $\varphi = 67.5°$. Intensities are given with respect to the maximum of the $E_{plasma}$ spectrum. Simulated spectra at $\varphi = 45°$ (not shown here) lead to a vanishing odd signal.

## 5. Discussion

The nonlinear peaks observed in the 2D spectrum in fig. 2c are combinations of the plasma frequency and the phonon frequency appearing at $(\nu_p, 0)$, $(\nu_p, 2\nu_p)$, $(\nu_{LO}, 0)$ and $(\nu_{LO}, 2\nu_p)$. Nonlinearities appearing at the plasma frequency $\Omega_p = \sqrt{e^2 n_e / \varepsilon_0 \varepsilon_\infty m^*}$ (with $n_e$ the electron density and $m^*$ the effective mass), especially the main peak $(\nu_p, 0)$, are due to small changes in the plasma frequency that can be attributed to two distinct effects taking place in the InSb system under high THz field excitation. The first effect is impact ionization, the electron-hole pair creation induced by energetic carriers scattering, which is known to occur in InSb under high THz electric field excitation [3]. Impact ionization induces an increase in the carrier density $n_e$, which leads to a change in plasma frequency. The second effect stems from non-parabolic features in the conduction band. When the conduction band electrons are driven by the THz electric fields under ballistic transport conditions, they dynamically explore the conduction band. If the band is non-parabolic, electrons will experience different effective masses for different $k$ positions and therefore changes in plasma frequency. We thus expect that both impact ionization and ballistic transport contribute to the nonlinear peak ($\nu_p$, 0).

Large-amplitude driving of electrons in a non-parabolic conduction band can also lead to nonlinear effects at harmonics of the plasma frequency, which result in the ($\nu_p$, $2\nu_p$) and ($\nu_{LO}$, $2\nu_p$) peaks in the 2D spectrum. We remark here that second order nonlinearities, which would appear at (0, $\nu_p$) and ($2\nu_p$, $\nu_p$) (corresponding to difference and sum frequency generation respectively, observed in [11]), are absent from the spectrum. This is expected given that the only second order nonlinear susceptibility tensor components $\chi^{(2)}_{ijk}$ of the zincblende structure of InSb that are non-zero are those with $i \neq j \neq k$. The second order polarization therefore vanishes at normal incidence on a (100)-surface and the observed peaks can be assigned to third or higher order nonlinearities.

We now focus on the nonlinearities observed at the LO phonon frequency $\nu_{LO}$, appearing as the ($\nu_{LO}$, 0) and ($\nu_{LO}$, $2\nu_p$) peaks in both the experimental and simulated 2D spectrum. A modification of the dielectric function due to a change in the plasma frequency leads to a redistribution in spectral weight which affects the nearby modes such as the LO-phonon mode, even in the absence of direct coupling. Then a THz field driven modification of the plasma frequency, through either carrier density or effective mass changes, leads to a change in phonon frequency [25] and then to the observed ($\nu_{LO}$, 0) and ($\nu_{LO}$, $2\nu_p$) peaks.

Impact ionization is an incoherent process due to the photo-excitation of electrons and evolves independently of the $E_{OR}$ phase. Ballistic transport, on the contrary, is a coherent process since the electrons are directly driven by both electric fields. These two effects can be partially differentiated by separating odd and even parity signals. In the even signal, the strong feature at ($\nu_p$, 0) contains incoherent changes due to impact ionization as well as a coherent contribution from the pump-probe effect of $E_{OR}$ on $E_{plasma}$. In figure 3, the peaks at ($\nu_{LO}$, 0) and ($\nu_{LO}$, $2\nu_p$), previously observed in figure 2c, are weak and so they are not discussed further. In contrast, the odd parity signal of fully cross-polarized beams contains only coherent process contributions. The peak ($\nu_p$, $\nu_p$) observed in the odd signal of fig. 3 for $\varphi = 22.5°$ and $\varphi = 67.5°$ emerges from a coherent process and is then a nonlinear signature of the non-parabolic conduction band experienced by the ballistically-driven electrons. Its intensity dependence with different $\varphi$ angles (it disappears for $\varphi = 45°$) leads us to conclude that this feature originates from the local band curvature experienced by the driven electrons for different directions within the conduction band. The weaker feature at frequency ($\nu_p$, 0) can be attributed to leakage from intense feature at ($\nu_p$, 0) in the even signal, due to non-perfect balancing of the polarity inversion.

The simulated spectra, from the last sets of fig. 3a and 3b, are in good agreement with the experimental results both in terms of intensity and angular dependence. In the simulated even signal, the ($\nu_p$, 0) amplitude is close to 50% of the maximum of $E_{plasma}$ spectrum, which is noticeably higher than the 30% found experimentally. The three secondary peaks observed experimentally, at ($\nu_p$, $2\nu_p$), ($\nu_{LO}$, 0) and ($\nu_{LO}$, $2\nu_p$) (see fig. 2c) are well reproduced in the simulated even signal but again with higher relative intensities than in the experiment. These discrepancies in intensity could be possibly due to impact ionization, which occurs in the experiments but is not included in the simulations. In addition, the simulation is based on a plane-wave approximation with homogeneous field-strength, while in the experiment, the sample is placed at the focus of the THz beam, where the peak field strength is only reached at the very center. In the simulated odd spectrum, the relative amplitude of the ($\nu_p$, $\nu_p$) peak is near 2%, comparable with the experiment. The simulated response for an angle $\varphi=45°$ (spectrum not shown here) shows no feature in the odd signal, also in good agreement with experimental observations.

The agreement between experimental data and simulations that include only ballistic transport indicate that ballistic transport dominates the THz nonlinear response in the first few picoseconds after excitation. This agreement is particularly good for the odd spectra where we were able to successfully isolate coherent ballistic transport signatures. We can then conclude that the odd spectral feature ($\nu_p$, $\nu_p$) is a manifestation of the band curvature symmetry experienced by the electrons, which are driven in different directions in the conduction band by the orthogonal THz electric fields.

We now further examine the local properties of the conduction band (band curvature) explored by the driven electrons. With FDTD calculations, one can follow step by step in the temporal domain the ballistic motion of electrons driven along the conduction band and the resulting nonlinearities. In figure 4a, we show the reconstructed electron trajectory along the conduction band from FDTD simulations evaluated at the sample surface, for $E_{plasma} = 35$ kV/cm, $E_{OR} = 65$ kV/cm and $\varphi = 0°$. This plot shows the fraction of the conduction band explored by the excited electrons. For these electric field amplitudes, the electrons span up to 5% of the Brillouin zone in $k_x$ and $k_y$, mostly along the direction of $E_{OR}$. Depending on $\varphi$ and on the polarity of $E_{OR}$, the electrons may encounter different lateral band curvatures, i.e. along $k_{OR}$, and orthogonal band curvatures i.e. along $k_{plasma}$. These differences in band curvature lead to different nonlinear responses. Figure 4b shows the evolution of the $(\nu_p, \nu_p)$ peak intensity from the odd signal with $\varphi$ angles between 0° and 90°. Results from FDTD simulations (blue line) are in good agreement with experimental values (red markers) extracted from the spectra in figures 3a and 3b. In particular, they show a strong dependence on the $\varphi$ angle, with maxima at 22.5° and 67.5° and minima at 0°, 45° and 90°.

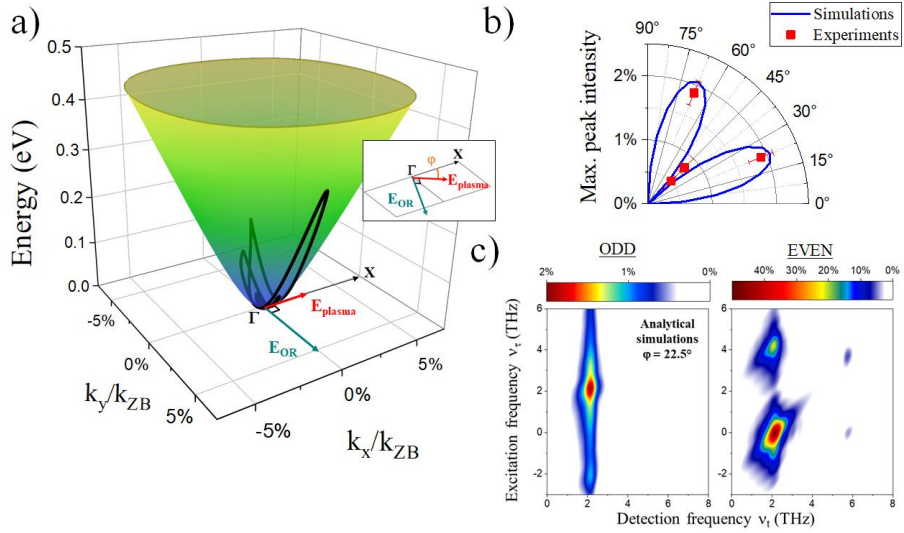

Fig. 4. a) Reconstructed electron trajectory along the conduction band for $E_{plasma}$ = 35kV/cm, $E_{OR}$ = 65kV/cm and $\varphi$ = 0°. Inset: $\varphi$ is defined as the angle between $E_{plasma}$ and $\Gamma\rightarrow X$, $E_{plasma}$ and $E_{OR}$ are kept orthogonal. b) Maximum of the $(\nu_p, \nu_p)$ peak intensity as a function of $\varphi$ angle, FTDT simulations using the InSb band structure calculated with a tight binding model are plotted as a blue line, experimental results for $\varphi$ = 22.5°, $\varphi$ = 45° and $\varphi$ = 67.5° are indicated as red markers. c) Simulated odd and even spectra using analytical model $E_{band3}$ for the conduction band (cf. text).

At this stage, it seems evident that the nonlinear signals in the simulations arise from a deviation of the conduction band dispersion from the parabolic free-energy model:

$$E_{band1} = ak_r^2 \qquad (2)$$

with $k_r^2 = k_x^2 + k_y^2$. This means that our technique can be used to investigate anharmonicity and anisotropy characteristics of the band curvature. In order to do so, we simulate the response for two different anharmonic conduction band models given by:

$$E_{band2} = ak_r^2 - bk_r^4 \qquad (3)$$

$$E_{band3} = ak_r^2 - bk_r^4 + c(k_x^4 + k_y^4) \qquad (4)$$

First, we will consider the fully parabolic band model given by eq. (2). Here our simulations show no nonlinearities. This is expected since the effective mass of a parabolic conduction band is constant for any *k*-vector (we recall that $m^* = \hbar^2(d^2E_{band}/dk^2)^{-1}$). For the model $E_{band2}$ with anharmonicity, we observe nonlinearities but only in the even channel. These are similar to features in the even spectrum previously simulated with the actual InSb band structure (fig. 3) and we retrieve the same features that combine plasma and phonon frequencies. Using model $E_{band3}$, with both anharmonicity and anisotropy, for $\varphi = 22.5°$, we obtain nonlinear contributions to both the odd and even signals, presented in figure 4c. Both odd and even spectra are similar to the experimental ones shown in figure 3. The dependence of the odd signal on the $\varphi$ angle was also verified for this model and ($\nu_P$, $\nu_P$) peak intensity dependence with the model $E_{band3}$ is very similar to the dependence with the InSb tight binding model (shown in figure 4b). Although these analytical models are too simple to accurately characterize the conduction band of InSb, they provide the right band curvature landscape within the k-range of the ballistic dynamics and explain qualitatively the experimentally measured nonlinearities. Specifically, we showed that the nonlinear signal of even parity is sensitive to the deviation from band harmonicity and the odd signal captures the anisotropy of the band structure. A more numerically efficient implementation of the FDTD calculation including higher order analytical models would enable approaching the conduction band curvature properties by fitting to the experimental data directly.

## 6. Conclusion

In conclusion, we studied the nonlinear dynamics of electrons in InSb by cross-polarized 2D THz spectroscopy highlighting the contribution of coherent ballistic transport in the first few picoseconds. The broadband nature of the plasma electric field enabled the observation of nonlinearities at combinations of plasma and phonon frequencies. With the support of FDTD-simulations, we described the temporal overlap regime and demonstrated that the nonlinear observations result from ballistic transport of electrons along an anharmonic and anisotropic conduction band. The spectra and their symmetries contain information regarding the local band curvature so that we can identify the symmetry of the valley in which the carriers reside. Our technique could in the future be applied to study a broad range of systems including the band structure of strongly nonparabolic electronic systems such as Dirac and Weyl semimetals in bulk, providing a possible complement to more direct measures of band structure that are typically, however, limited to near-surface states.

## 7. Funding and acknowledgements

This work was supported by the NCCR Molecular Ultrafast Science and Technology (NCCR MUST), a research instrument of the Swiss National Science Foundation (SNSF). E.A. acknowledges support from the ETH Zurich Postdoctoral Fellowship Program and from the Marie Curie Actions for People COFUND Program. We acknowledge Georg Winkler and Quansheng Wu, previously from the Institute for Theoretical Physics and Station Q at ETH Zurich, for the band structure calculations.